# Structure and Properties of Rare earth Neodymium Zinc Titanate

Kouros Khamoushi
Faculty of Automation, Mechanical and Materials Engineering, Tampere University of Technology,
Korkeankoulunkatu 6 33720 Tampere Finland, Kouros.khamoushi@tut.fi

**Abstract**

The dielectric properties and phase grouping of Rare earth Neodymium Zinc Titanate (NZT) was investigated in this research. The result shows that it is distrustful to be a stable perovskite structure, in fact something comparable to Ilmenite structure, however the further research shows that the monoclinic structure can be purposed for NZT. The Modelling and simulation were used in this research to define the atomic position and crystal structure of NZT. These compositions have promising dielectric properties and can be used in microwave telecommunications.

.
## 1. Introduction

The extend to reduce the size, weight and cost of microelectronic devices principles for tuning $\tau_f$ in complex perovskite have shown that $\tau_\varepsilon$ in complex perovskite materials is fundamentally related to the start and degree of octahedral tilting. $La(Mg_{1/2}Ti_{1/2})O_3$ (LMT) is a perovskite similar to $Nd(Mg_{1/2}Ti_{1/2})O_3$ NMT which forms in the monoclinic system, its tolerance factor is t = 0.9521, which indicates the presence of both in-phase and anti-phase tilting of oxygen octahedra[1], Systematic cation displacement was also detected by the appearance of 1/2(210) and 1/2(111) reflections and caused by the B-site cation ordering[2]. Both result in doubled unit cells however, cation ordering also reduces the space available in which the A-site species can rattle. In this study, the structure and microwave dielectric properties of $Nd(Zn_{1/2}Ti_{1/2})O_3$ NZT were investigated which is isostructural to LMT and NMT.

## 1.1. Miniaturization of Resonators

The system on package can give flexibility to the front-end module by integration in all functional blocks using multilayer processes and novel interconnection methods. In these system a multilayer blocks of electrical components are used to be able to reduce the size of components. The utilization of small sizes of resonators in communication and low frequency such as cellular communication and Wireless fidelity (WiFi) is very difficult tasks to fulfil.

The size reduction of resonators in a package has also advantages such as thermal loss reductions. The integration of resonators has also connection with Bandwidth which should be narrow and must have high dielectric or relative constant. The interference of resonators and the rest of radio frequency blocks must have high modulation and be in phase. When the inductors and capacitance size are changed (in this case becomes smaller) the resonators size also must be changed, because these 3 components are connected to each other.

The gain of resonators depends on its electrical size. In Patch resonators the wave length should be in the order of $\lambda/2$
in which $\lambda$ is the wave length. For example, at 2 GHz, the electrical wave length in FR4 dielectric (dielectric constant = 4). This means that the dielectric is equal to 4 and wave length is equal to 74 mm in practical it will be 1/2 of that it means 37 mm. There are many techniques in which the RF front to ends which is coming from resonators and goes to transreceiver can have a size of $5\times 5$ mm for example by multiple transmit receiver this is possible, which will be 7 times smaller than resonators itself. Therefore, resonators can be a RF front end for reduction of size.

There are 3 important facts in resonators which must fulfil. (1). Resonators *should be physically small*. (2). *It should have a narrow bandwidth*. (3). *It should have a good gain*. The size of resonators is function of its *"materials"* and properties and by following equation it can be calculated.

$$L = \frac{\lambda}{\sqrt{\mu_r \times \varepsilon_r}} \text{ in which:} \qquad (1)$$

L= physical size of antenna
$\lambda$ =resonators electrical size in air
$\mu_r$ = relative permeability
$\varepsilon_r$ = relative permittivity

If there is a dielectric materials equal to 9, the size of resonators will be 1/3 of that when $\mu_r$ =1 for air. The calculation of band width is occurring by $BW = \frac{F_{high}-F_{low}}{F_{center}} \times 100$ in which $F_{high}$ is the highest frequency and $F_{low}$ is lowest frequency and the $F_{center}$ is centre frequency. There is *Multiband resonators, conformal resonators* and resonators on Magneto dielectric substrate. Briefly the resonators with small size must have materials with *high permeability, high dielectric constant* and high permittivity property. This materials operation is to prevent the parasitic substrate mode which is reducing the gain of resonator[3].

Gain is constantly measured with reference to standard resonator such as an isotropic or dipole resonator. When reference is an isotropic a resonator, the units used are dBi. Since an resonator only redistributes power that oftentimes is non-uniform in space, the gain changes as a function of direction and therefore has both positive and negative values. For consumer applications, an omnidirectional radiation pattern is often preferred.

The gain of an resonator is a function of many factors, the most important of which is resonator matching. The transmission line that feeds the resonator has to be well matched to the resonator input to ensure a maximum transfer of power. This can be very challenging task in resonator miniaturization.

To summarize, the resonator with small size needs materials with high permeability and permittivity. However materials with such kind of properties generate parasitic substrate modes that deteriorate the gain of resonators with higher gain necessitate matching the resonator input port for maximum transfer of energy from transmission line to the resonator. With high permittivity and high permeability materials, this becomes difficult since very narrow traces are required to obtain an impedance of 50 $\Omega$, with such a conflicting requirements, the design of resonators can be very tricky.

A variety of advances have been suggested by a number of researcher for miniaturization of these resonators such as combining multiple frequency bands into a single resonator elements, use of magneto dielectric materials, conformal resonators and inclusion of electromagnetic bandgap structures to improve performance.

Another important factor that influence and require special attention while integrating the resonator in the module is suppression of parasitic backside radiation or crosstalk, which can couple to sensitive RF blocks in the substrate. This is measured as the front to back ratio, which is the ratio of maximum directivity of resonator to its directivity in the backward direction.

However three important resonator technologies for WiFi applications will be described here for miniaturizing the resonator size and integrating it with the rest of the RF front-end module.

These three important resonators are: multiband resonators, conformal resonators, and resonators on magneto-dielectric substrates. Since several communication standards are integrated into the RF front end, there is a requirement for resonators to support several frequencies.

Along with passing the enquired frequencies, these resonators should also be minimize, interference by suppressing the adjacent frequency bands. To minimize, a multiband resonator instead of multiple single-band resonators can be constructed by controlling the length of the resonator elements. In system on package it is possible to combine rigid and flexible substrate. This phenomenon can be used to embed radio frequency front to end in the rigid part of module. Resonator which is patterned on the flexible substrate can be folded or prepared to conform the rigid section. This method is reducing the size of module containing the integrated antenna.

**1.2. Waveguide and modes**

The solution of Maxwell's equations with the boundary conditions for tangential conditions inside and outside of the resonators can lead to an eigenvalue problem; this problem consists of differential equations called resonant modes, which are eigenfunctions.

For each eigenfunction, there is the corresponding eigenvalue, which is the resonant frequency. In the propagation of electromagnetic wave, many resonance modes exist in resonant chamber.

For each eigenfunction, there is a corresponding eigenvalue; which is the resonant frequency. The *modes* are various reflected waves that interact with each other to produce an infinite number of discrete patterns. When two open sides along the length of a waveguide are closed by two perfectly conducting planes, the waveguide is called cavity resonators[4].

In the propagation of an electromagnetic wave, many resonance modes exist in the resonant chamber. They are longitudinal and perpendicular to the direction of the propagation.

If the magnetic and electric fields are existing, and have a transverse electromagnetic mode, they are shown by TEM mode; if the electric field is transmitted but the magnetic field is zero; then the mode is called transverse magnetic, or TM**.**

If the electric field is zero, and only longitudinal, (Z direction) propagation exists, then the mode is called transverse electric, or TE mode. The discussion of all these modes is outside of the scope of this research. The work here will be concerned with TE modes. In these studies the resonant frequencies of the $TE_{01\delta}$ mode was measured at each temperature. The waveguide transmission characterized by two significant conditions:

1) The waveguide has a minimum frequency below which it cannot transmit the wave
2) There is always a component of Electric and Magnetic field along the direction of field.

The electromagnetic wave can enter in one end and can be received at the other end. The ionosphere and the earth act as a wave guide for radio wave.

There are Rectangular mode and Circular modes designed by $TE_{mn}$ and $TM_{mn}$. The subscript (m) means the number of half sinusoidal variation of field along dimension (a), while subscript (n) refers to the field dimension along (b). However in Circular modes, there are two kinds of modes $TE_{mn}$ modes and $TM_{mn}$ modes. The subscript (m) means number of full sinusoidal variations of field along the circumference and the (n) subscript refer to the number of half sinusoidal variations of field along radius $TE_{01}$.

Furthermore, there are two more subscripts (p) and (q) designates the number of half wavelength along guide length (*l*). Basically, the $TE_{mnp}$ and $TM_{mnp}$ modes have application in rectangular cavities and simultaneously $TE_{mnq}$ and $TM_{mnq}$ modes, are used in circular cavities. For example let consider a rectangular cavity of sides a, b and c which have perfectly reflecting wall. If the plane wave in space is defined by vector k, normal to the plane of wave with three components called $k_1$, $k_2$, and $k_3$ along X, Y and Z axes. When a wave produce inside of the cavity, it is reflecting in all directions and set of waves such as: $+k_1$, $+k_2$, $+k_3$; and $-k_1$, $-k_2$, $-k_3$ will produced, these wave will be constructive if they are in phase with standing wave inside of the cavity. The k wave in direction of a, b, and c and can be defined as:
$k_1 = \pi (m/a)$;   $k_2 = \pi(n/b)$;   $k_3 = \pi (l/c)$;
m, n and l bare integers equal to number of wave in a, b, and c directions, since

$$k = \sqrt{k_1^2 + k_2^2 + k_3^2} \qquad (2)$$

$$k = \pi \sqrt{\left(\frac{m}{a}\right)^2 + \left(\frac{n}{b}\right)^2 + \left(\frac{l}{c}\right)^2} \qquad (3)$$

The frequency of sanding wave

$$f_{mnl} = \frac{c}{2}\sqrt{\left(\frac{m_1}{a}\right)^2 + \left(\frac{n_2}{b}\right)^2 + \left(\frac{l_3}{c}\right)^2} \qquad (4)$$

where c is equal to velocity of electromagnetic wave, f is frequency. A cavity such as the one shown in **Figure 1**, will resonates by supporting standing wave for the frequency given by equation 4. In other conditions, for example, when the cylindrical or the spherical cavities are used, the mathematical treatment is more difficult. The resonators cavities for electromagnetic wave must have the walls which are made from a good conductor material, because the walls must have the best reflecting properties[5].

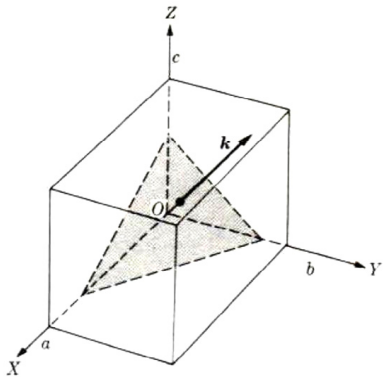

**Figure 1**. A cavity sustaining standing wave[5].

The Cylindrical closed cavity was used in this research work to examine the samples; therefore this will be described here briefly. These cavities are easy to construct and are greatly used for measuring quality factor and temperature coefficient of resonant frequency. The cavity modes are selected for measuring, the $TE_{mnp}$ and $TM_{mnp}$.

Where (q) refers the number of $\frac{\lambda_g}{2}$ along the length that are shown in **(Figure 2)** [6].

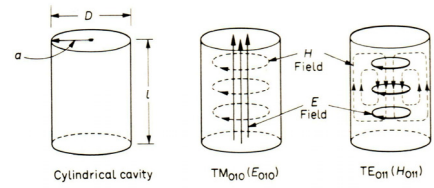

Figure 2. The TM and TE different modes for wave-meters [6].

Let suppose the ($\lambda$) is the resonant wavelength of cavity and (l) is the length and (a) is radius shown in Figure 2, then for TE modes

$$\frac{1}{\lambda^2} = \frac{1}{\lambda_c^2} + \frac{1}{\lambda_g^2} \qquad (5)$$

$$l = \frac{q \lambda_g}{2} \qquad (6)$$

q = 0, 1, 2 etc and ($\lambda_c$) is cut-off wavelength (minimum frequency below which waveguide will not transmit wave) and $\lambda_c = 2a$; and ($\lambda$) is space wavelength; and ($\lambda_g$) is Cylindrical guide wavelength. $\lambda_g > \lambda$. Since $\lambda_c = \frac{2\pi}{k_c}$ and $\lambda_g = \frac{p}{2l}$ after substitute in equation 20 it will give;

$$\frac{1}{\lambda^2} = \left(\frac{k_c}{2\pi}\right)^2 + \left(\frac{q}{2l}\right)^2 \qquad (7)$$

The $k_c = \frac{p'_{mn}}{a}$ where $P'_{mn}$ is the nth roots of the equation $J'_m(k_c a) = 0$ then

$$\frac{1}{\lambda^2} = \left(\frac{P'_{mn}}{2\pi a}\right)^2 + \left(\frac{q}{2l}\right)^2$$

$$\frac{1}{\lambda^2} = \left(\frac{P'_{mn}}{\pi D}\right)^2 + \left(\frac{q}{2l}\right)^2$$

$$\lambda = \frac{1}{\sqrt{\left(\frac{P'_{mn}}{\pi D}\right)^2 + \left(\frac{q}{2l}\right)^2}} \quad \text{(TE modes)}$$

This is similar to TM but $P'_{mn}$ will be replaced by $P_{mn}$ in case of TM modes and it will be

$$\lambda = \frac{1}{\sqrt{\left(\frac{P_{mn}}{\pi D}\right)^2 + \left(\frac{q}{2l}\right)^2}}$$

**1.3. The Differential Scanning Calorimetry**

The thermal analysis such as Differential Scanning Calorimetry (DSC) is used to determine the thermodynamic changes of various powders of ceramic materials or reaction changes between the materials and the atmosphere. The device (Perkin Elmer TGA7) includes two pans; the sample pan and reference pan. The powder located in the sample pan and the reference pan was empty.

Each pan stands on top of the heaters, the computer turns on the heaters, and it was heating the two pans at a specific rate of 10 $^oC$ per minute. Samples were heated over a programmed temperature range of (30-900 $^oC$).

The computer was used to make sure that the heating rate stays exactly the same throughout the experiment; in other words, to keep the temperature equal in the reference and sample pan. The heater underneath the sample pan has to work harder than the heater under the reference pan; because it needs to generate more heat. This technique is called DSC; when it is measuring this extra heat based on the change in enthalpy.

The changes can be an exothermic or an endothermic transition reaction in the sample, as a function of time and temperature. The DSC and its apparatus are shown schematically in **Figure 3**.

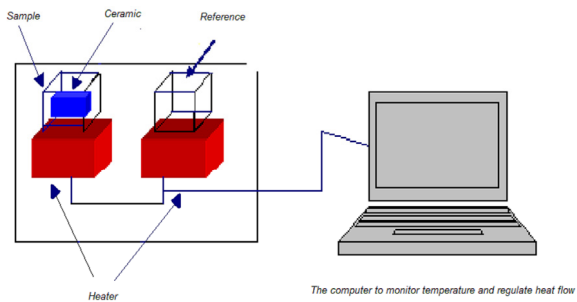

**Figure 3.** A Schematic Differential Scanning Calorimetry Measurement.

### 1.4. DSC Measurement of Neodymium Hydroxide

The thermal analysis by DSC shown graphically in (**Figure 4**) at 400 $^oC$, the Neodymium Hydroxide which was result from mixing of hexagonal $Nd_2O_3$ in water, it has lost two third of its water, and probably becomes NdOOH according to following formula.

$$Nd(OH)_3 \xrightarrow{Heat} H_2O + NdOOH$$

$$Nd(OH)_3 \xrightarrow{371\ ^oC} H_2O + NdOOH$$

$$Nd(OH)_3 \xrightarrow{371\ ^oC} NdOOH$$

$$NdOOH \xrightarrow{550\ ^oC} C\text{-}Nd_2O_3$$

Above 550$^o$C the NdOOH dries further to become cubic $Nd_2O_3$, a much less hydratable polymorph of the oxide; from ~800 to ~1000 $^oC$ it reverts back to hexagonal $Nd_2O_3$. The XRD of powder samples of Neodymium Hydroxide at 400 °C and 600 °C were preformed. The Crystallographica Search Match program was used to define the crystal structure of these samples.

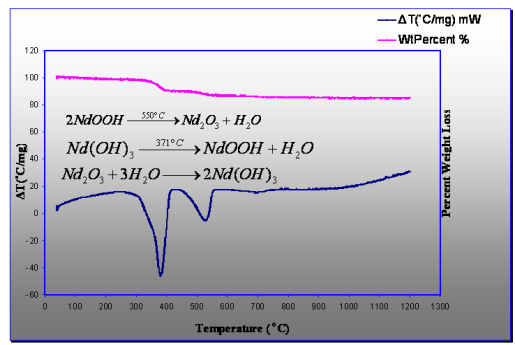

**Figure 4.** Thermal analysis of $Nd(OH)_3$.

In the thermal analysis of Hexagonal $Nd(OH)_3$, first hexagonal $Nd_2O_3$ powder was mixed with water to produce $Nd(OH)_3$. Because the $Nd_2O_3$ absorbs water from the air, powder was purposely hydrated first in distilled water to produce $Nd(OH)_3$; then it was used as a starting powder for the preparation of NZT. The XRD test was used to show that. The crystal structure above 370 °C was transited to monoclinic. Above 600 °C it reverts back to a hexagonal crystal structure again. Around 400 °C the strongest peaks of the XRD sample powder matched the monoclinic crystal structures. However above 550 °C powder dehydroxylated further again to a cubic crystal structure.

The reflections marked $\alpha_1$ and $\alpha_2$ in **Figure 5** correspond to two different sets of {*odd, odd, odd*} reflections, which are associated with antiphase tilting of oxygen octahedral along at least two axes. The peaks labelled $\beta_1$, $\beta_2$, correspond to antiparallel displacement of A-site cation, $Nd^{+3}$.

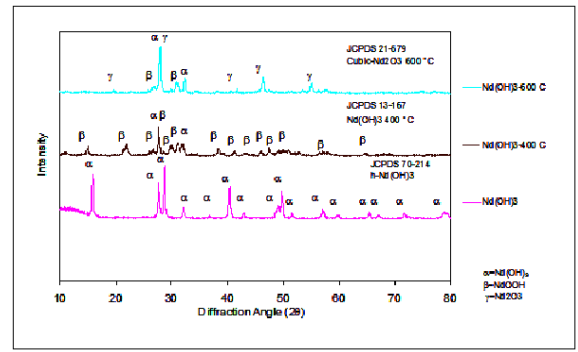

**Figure 5.** XRD of $Nd(OH)_3$ (Hexagonal); and product of calcination of $Nd(OH)_3$ at 400 °C (monoclinic) and $Nd(OH)_3$ at 600 °C is cubic.

### 2. Vibrational spectroscopy

The samples of Neodymium Zinc Titanium Oxide (NZT) were examined by Raman spectroscopy therefore, in this section basic information of Raman Spectroscopy will be described. Vibrational spectroscopy, Raman and infrared, are derived from the interaction of electromagnetic radiation with material. These tests are usually non-destructive, the plan with the research is to investigate the molecular vibrations, which are not observed with x-ray or other methods. Theses molecular vibration can be observed as bands in the vibrational spectrum. Each structural configuration and molecule will have a certain set of bands, which are result of their intensities, frequencies and shapes. The information about the local coordination of the atom can be achieved by analyzing the sample properties. The frequency of a band depends on the reduced mass of the travelling unit and it is normally around ~50 to 4000 cm$^{-1}$.

The vibrational frequency depends directly on reduced mass therefore, the smaller reduced mass cause higher vibrational frequency and vice versa.

Raman or infra red spectroscopy can be used for testing a sample, however it is depending on the nature of the molecular vibration.

The physical principle behind Raman and infrared spectroscopy is very different despite the fact that, Raman spectroscopy is a scattering technique where the intensity of the bands is proportional to the concentration, and the scattering cross-section of the vibrating units. Infrared spectroscopy is an absorbance technique where the intensity of the bands is proportional to the concentration, and absorption coefficient of the vibrating units.

The time needed to record a spectrum is of the order of seconds to tens of minutes. However, to increase the signal-to-noise ratio, several accumulations are expected[7]. Raman spectra arise from monochromatic laser light impose on a specimen, this light is then elastically and inelastically scattered **(Figure 6)**. Elastic scattering, Rayleigh scattering, has the highest probability. The Raman spectra are result of inelastic scattering of incident energy from a laser source as a function of molecular energy. It is commonly due to photons that combine to molecular vibrations or phonons in the material. The Raman scattered light can be of higher energy, anti-Stokes scattering, and lower energy, Stokes scattering, than the incident light. Stokes scattering occurs with the highest probability, therefore one usually record the Stokes bands in the Raman spectrum[8].

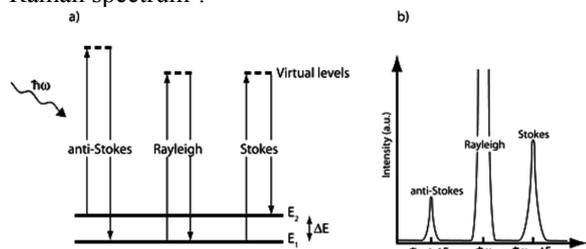

**Figure 6.** a) Schematic picture of the Raman process in a two-level system, and b) the resulting Raman spectrum. $\omega_L$ = Laser frequency; $\omega_S$ = Raman frequency shift; Q = normal coordination of vibration; α = Molecular polarizability; μ = molecular dipole moment; $n_s$ = Bode-Einstein factor; I= Intensity; λ = wavelength[8].

A two stage system are shown energy level schematically in Figure 6, as shown the scattered light is result of anti-Stokes, Stokes and Rayleigh, the transition level is result of the changes between the two vibrational states and occurs through a forbidden virtual level. Raman scattering is not occur because of direct change. The terms for the Stokes and anti-Stokes intensities can be describe as

$$I_{stokes} \propto (\omega_L - \omega_S)^4 \frac{h/2\pi}{2\omega_S} \left(\frac{\partial \alpha}{\partial Q_S}\right)^2 (n_S + 1)$$
(8)

$$I_{anti-stokes} \propto (\omega_L - \omega_S)^4 \frac{h/2\pi}{2\omega_S} \left(\frac{\partial \alpha}{\partial Q_S}\right)^2 n_S$$
(9)

The intensity expressions above gives the important results. If the specific molecular vibration adjust the polarisability of the molecules then Raman scattering will take place and it means

$$\left(\frac{\partial \alpha}{\partial Q_S}\right)^2 \neq 0$$

## 2.1. Infrared spectroscopy

The differences between Infrared and Raman spectroscopy is coming from the point that in Infrared spectroscopy molecular vibration adjusting the dipole moment while in Raman spectroscopy molecular vibration adjusting the polarisability of molecules Molecular vibration that adjusting the dipole moment can be occur if following condition is fulfilled:

$$\left(\frac{\partial \mu}{\partial Q_S}\right)^2 \neq 0$$

The molecules such as diatomic homonuclear material does not have dipole moment , therefore it does not increase any band energy in Infrared spectra. When incoming light impinging on the sample the interaction between photons and materials may cause the absorption of photons if their frequency of photons is accurately matches the frequency of vibration in materials. This frequency is absorbing frequency of photons. The Infrared spectroscopy is illustrated in **Figure 7**.

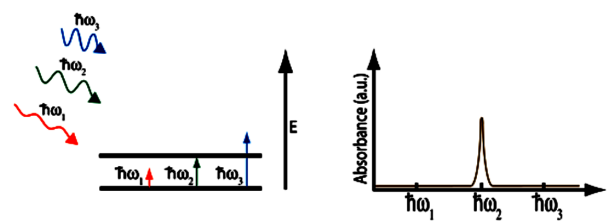

**Figure 7.** The principle of infrared absorption. a) Photons with energies $h\omega_1$, $h\omega_2$, and $h\omega_3$ hit a two-level system. Only $h\omega_2$ has the same energy difference between the two vibrational states, and is therefore

absorbed. b) The resulting infrared absorbance spectrum[8].

**2.2. Use of symmetry analysis to vibrational spectroscopy**

The aim of this section is to give a very short introduction to symmetry analysis and group theory, which can be used to determine the number and symmetry of vibrational modes of a particular molecule in the samples.

**Symmetry operators and symmetry elements**
The symmetry of molecules take an important place for defining the crystal structure of materials and their space group. Angular momentum is a part of group theory, as well as are the properties of harmonic oscillator.
In order to make use of the symmetry of molecules and crystals for vibrational spectroscopy, the symmetry properties have to be defined conveniently. However, here symmetry operators will be define, which applied to a molecule, bring it into coincidence with itself. For molecules there are five different symmetry operators, which are defined as below.

E    identity operator, the act of doing nothing. The corresponding symmetry elements is object itself.
I    An inversion operator, reflection at a centre of symmetry. This operation taking each point of an object through its centre and out to an equal distance on the other side.

$C_n^k$   describes an n-fold rotation operator; in which $k$ is consecutive rotations by an angle $2\pi/n$, where $n$ is the order of the axis of symmetry rotation.

σ    A reflection operator, reflection on a plane, in a mirror plane. In the principal axis of symmetry, it is called vertical plane and denoted $\sigma_v$. If the principal axis is perpendicular to the mirror plane, then the later symmetry element is called a horizontal plane and denoted $\sigma_h$. A dihedral plane denoted $\sigma_d$ it is a vertical plane that bisects the angle between two $C_2$.

$S_n^k$   n-fold improper rotation -reflection operator; defining k successive rotation about an $n$-fold rotation-reflection axis: rotation by an angle of $2\pi/n$ followed by reflection on a plane perpendicular to the axis.

**2.3. Symmetry elements**

To each symmetry operation there are corresponding a **symmetry elements** such as the point, line, or plane with respect to which operation is carried out. For example a rotation carried out with respect to a line called axis of symmetry, and a refection carried out with respect to a plane called mirror plane. If the translational symmetry operation is ignored then there are 5 types of symmetry operations which leaves the object to be unchanged. These are

| | |
|---|---|
| E | 1 |
| I | $\bar{1}$ |
| $C_n^k$ | 2,3,4. |
| $S_n^k$ | 2,3,4. |

The main axis are those with highest order, it is by definition oriented vertically around (z) direction.
The axis with the highest order n is called the main axis, and is by definition oriented in the vertical direction (*z*). For example, let x-y plane to be is horizontal level and perpendicular to vertical z axis direction denoted by $\sigma_h$. A plane, which includes the main axis is a vertical plane $\sigma_v$.
The crystal structure of materials according to its symmetry is depending on all its symmetry operations, and then put down a label based on the list of those operations simply the list of symmetry operation is used to identify the **point group** of the crystal. The point group indicates that the operation corresponding to symmetry elements that intersects at a fix point or origin.
It should be noted that a point group plus translation symmetry called **space group.**
The 32 point group are allowed in crystals, for example the point group **C2** includes symmetry operators of E, and $C_2$.
The symmetry operators of the 32 point groups of crystals can be combined with the translations unit and creating 230 three-dimensional **space groups**. However, this is out of scope of this studies.
There are two kind of molecular vibration, it can be symmetric or anti-symmetric depends on symmetry operation of its point group. When a molecule defined by its $C_n$ or $S_n$ and n > 2 it means two or more vibrations have the identical frequency.

The symmetry operation can be described by matrices. Let the symmetry operator G produces the vector [x+y+z] into [x'+ y'+z']. This could be written as G[x+y+z]= [x'+ y'+z'], or as follows:

$$\begin{bmatrix} x' \\ y' \\ z' \end{bmatrix} = \begin{bmatrix} C_{xx} & C_{xy} & C_{xz} \\ C_{yx} & C_{yy} & C_{yz} \\ C_{zx} & C_{zy} & C_{zz} \end{bmatrix} \cdot \begin{bmatrix} x \\ y \\ z \end{bmatrix}$$

The positive direction of individual old as well as new axes are defined by **G** matrix which stand for matrix with cosines angles between positive direction of axis. The **G** operator represent the transformation matrix. All operators group are represented by matrices.
The $\Gamma_{reduced}$ is stand for reduced representation. A representation $\Gamma$ of a group consist of set of numbers or matrices which can be assigned to the element of group in such a way that the same group multiplication relationships are assured.

## 2.4. Irreducible representation and character tables

The degree of freedom for the molecules containing n atoms is 3n. It is necessary to apply all point group of molecule to each atoms to be able to classify them. This classification results in a representation by a group (g) square 3n×3n matrices. However, there is an operation which reduces these matrices to the matrices of same dimension and it is called similarity transformation. These are lower order sub-matrices along the diagonal. The symmetry operator is represented by trace of matrix called character, and it is denoted by $\chi$.

If transformation matrix does not exist that can reduce the representation, therefore it is called complete, the irreducible representation are denoted by $\Gamma_1$, $\Gamma_2$, $\Gamma_3$…..These are sub matrices

The completely reduced representation is result of the sum of irreducible representation of **ni** Multiply by individual reduced representation, that is

$\Gamma_{reduced} = n_1\Gamma_1 \oplus n_2\Gamma_1 ….$

The lists of all possible irreducible representations of point group are illustrated in table

The character table for point group $C_{2v}$ is described below. The top row shows the lists of operators. The irreducible representation are shown in left column, with their name of vibrational type.

In the right side of table and in the last column after A1 there is z, $x^2$, $y^2$, $z^2$, these variable are representing A1 symmetry. In the similar way A2 symmetry has a connection with xy, and R . Here the R stand for rotation and subscript x shows rotation around the x-axis.

The character table row illustrate its symmetry

The characters of the irreducible representation symmetry properties are included in rows of Character **Table 1.**

By the characters of the irreducible representations in the row of a character table, its symmetry properties will be defined.

**Table 1.** Character Table

| $C_{2v}$ | E | $C_2$ | $\sigma_v'$ | $\sigma_v''$ | |
|---|---|---|---|---|---|
| A1 | 1 | 1 | 1 | 1 | z, $x^2$, $y^2$ |
| A2 | 1 | 1 | -1 | -1 | $R_z$, xy |
| B1 | 1 | -1 | -1 | -1 | x, $R_y$, xz |
| B2 | 1 | -1 | -1 | 1 | y, $R_x$, yz |

The group theory has one important and it is connected to orthogonality theorem, it means that number of ($n_i$) vibrations which is belong to specific symmetry types (i) can be calculated for example A1. These are described in next equation

$$n_i = \frac{1}{g}\sum_g \chi^{(i)}(R)\chi^{red}(R)$$

The number of symmetry operators of the group is equal to g. The symmetry operator is (R) and $\chi^{(red)}$ is its character of reducible representation and for the symmetry operator $R$ the $\chi^{(i)}$ is the character of the $i$:th irreducible representation.

## 2.5. Electrical Measurement

The electrical measurement including the quality factor (Q) and the temperature coefficient of the resonant frequency ($\tau_f$) were made at London South Bank University, using a network analyzer (Hewelt Packard-8719C 50MHz-13.5 GHz). $\tau_f$ was measured for LZT and NZT at Filtronic Comtek. The quality factor Q was tested by using a cylindrical resonance cavity made of high conductivity metal, with an interior approximately 3-5 times larger than the dimensions of the test sample. The test sample was placed inside the cavity upon a low loss, low dielectric constant support, and inductive coupling to the resonator sample was achieved; via a coupling loop or a bent probe. The transmission characteristics of the $TE_{01\delta}$ resonant mode were measured, and the quality factor was calculated using the formula, $(f_o/\Delta f)/(1-10^{-(IL/20)})$ where $f_o$ is the resonant frequency; $\Delta f$ is the 3dB bandwidth; and IL is the insertion loss expressed in dB. The quality factor at very low frequencies (~1KHz-1MHz) may also be measured using various commercially available capacitors or L.C.R. meters and fixtures, which give a direct reading of $\tan\delta$.

This technique; however, often results in quality factor data which are extremely inaccurate and potentially misleading, unless all variables are properly accounted, especially for high quality factor materials. It is best used for relative comparisons of similar samples. The device (Hakki and Colman) for measuring the temperature coefficient factor and the quality factor are shown in Figure 8.

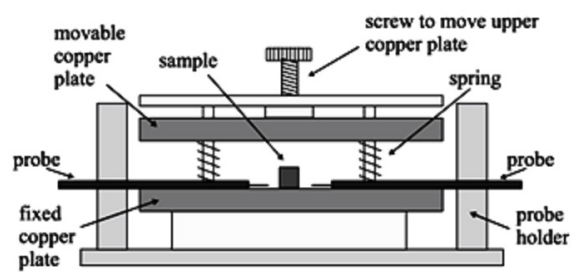

**Figure 8.** A Schematic device for measuring the temperature coefficient factor and the quality factor with Hakki-Coleman configuration[9].

Before sintering, the calcined powder was pressed with a cylindrical shape of 10 mm in diameter and 4 mm in thickness; at an isostatic pressure of 125 Mpa. The dielectric constant and unloaded Q at a microwave frequency were measured using a parallel-plate method, with a net network analyzer (Hewelt pakard-8719 C). The temperature coefficient of the resonant frequency was measured at 4-11 GHz in a temperature range from -23 +37 °C. The thermal stability of a resonator is defined by the temperature coefficient of the resonant frequency. It indicates graduate changes of the resonant frequency with the changing temperature. Our aim is to keep ($\tau_f$ or TCF) as near to zero as possible. Therefore it is important that the frequency of the microwave device does not drift. Table 2. shows the temperature coefficient of the resonant frequency of NZT.

**Table 2**. Lattice parameters, theoretical density and temperature coefficient,
permittivity, the quality factor of LZT and NZT.

| Reso nator | Lattice Para. (a) nm | Lattice Para. (b) nm | Lattice Para. (c) nm | $\rho_{th}$ g/cm$^3$ | $\tau_f$ M K$^{-1}$ | $\varepsilon_r$ | Q×f |
|---|---|---|---|---|---|---|---|
| NZT | 0.7869 | 0.7564 | 0.8049 | 6.901 | -47 | 36 | 9800 |

## 3. Experimental procedure

In this work, conventional mixed oxide powder processing techniques were used. Starting materials included, $Nd_2O_3$ (99.9% Meldform Rare Earths, U.K.), $TiO_2$ (99.8% Alfa Aesar, U.K.), and ZnO (99.9% Elementies Specialties, U.K.),$(MgCO_3)_4Mg(OH)_2 5H_2O$ (99% Aldrich Chemical company, Inc, USA), $Nd_2O_3$ powder are notoriously hygroscopic and readily hydrate in the atmosphere if unprotected. Consequently, many groups involved in the mixed-oxide processing of rare-earth-containing ceramics have tried milling in various solvents, including ethanol,[9,10,11] methanol,[12] and acetone,[13,14] or even dry;[15] although the use of distilled[16,17] or deionised[18] water is not unprecedented. Some researchers[19] have even adopted the practise of storing these powders at 800°C until needed in order to keep them dry; however, as this procedure is expensive, and the use of solvents has unpleasant environmental implications, another method was tried in this study. For NZT , the rare-earth oxide was first purposely hydrated in distilled water to form $Nd(OH)_3$. These hydrates were then used in the subsequent processing procedure,
which involved milling stoichiometric amounts of powders together in a porcelain mill pot partly filled

All the samples show a structure similar to perovskite structure. Previously a high degree of octahedral tilting crystal structure was suggested for NZT[20,21].The result of this investigation shows that, the crystal structure similar to ilmenite structure can be proposed for NZT. However, other research suggested a monoclinic structure[22]. The tolerance factor of NZT is t=0.916 while tolerance factor of $ABO_3$-type ilmenite[23] is with $ZrO_2$ media and distilled water for four hours. A small amount (1wt%) of Dispex A40 (Allied Colloids, Bradford, U.K.) was added as a deflocculant. The slurries were then dried overnight at 80°C. Dried powders were subsequently granulated with a mortar and pestle and sieved to less than 250 μm. Calcination was achieved using a two-stage process. First, powder was heated to 650 °C for 2 hours in an open $Al_2O_3$ crucible to ensure dehydroxylation of the $Nd(OH)_3$. The completion of the dehydroxylation reaction was monitored by measuring the weight loss at this stage of the process. Second, this same powder was gently mixed by hand, a lid was placed over the crucible, and it was re-heated to 1400ºC for 2 hours. Afterwards, the powder was re-milled for a further four hours with 2wt% PEG 1500 (Whyte Chemicals, London) being added in aqueous solution 15 min before completion. These slurries were then dried and granulated as above and subsequently pressed (125 MPa) into cylindrical pellets 10 mm in diameter and 3 mm thick. Sintering was conducted in closed alumina boats for 6 hours at temperatures ranging from 1450°C to 1675.

The densities of NZT as a function of sintering temperature were investigated in this study. The calcinations temperature for NZT was 1200 ºC and sintering temperature 1450 ºC. **Figure 9** illustrates the relative density of NZT as a function of sintering temperature. The densities of NZT first increase with temperature up to maximum of 6.67% g/cm$^3$ of theoretical density at 1450 °C and then slowly decrease at 1500 °C. The decrease in density at higher sintering temperature is thought to be due to an increase in amount of ZnO volatilization, which increases rapidly at temperature above 1500 °C for NZT.

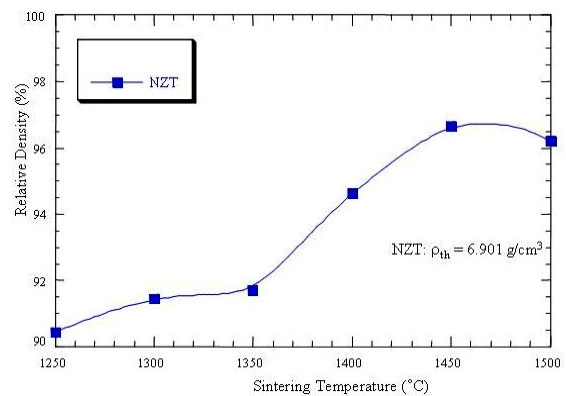

**Figure 9.** Relative density of NZT pellets after sintering for 6 hours at 1250-1500 ºC.

## 4. Results and Discussion

t=0.80. These results shows that NZT deviated from perovskite about 8.4% while deviation for ilmenite is 20%. Probably, there is some similarity between the structure of ilmenite and NZT, but result shows that NZT crystal structure is closer to perovskite in contrast to ilmenite.
The XRD diffraction traces of NZT from 1250 °C to 1500 °C are shown in **Figure 10**. These samples indexed using pseudocubic lattice parameter a=0.3939 nm, there was slight shift and broadening in position of peaks which indicate the existence of monoclinic crystal system. The x-ray diffraction alone is not sufficient to determine the symmetry of tilted perovskite. Therefore prediction of crystal structure method used for additional aid. Bond valence methods were used in the theoretical calculations of interatomic distances. The method is based on the modified second rule of Pauling[24] and Baur. The calculation showed that the bond length of Neodymium to oxygen (Nd-O) is equal to 0.3232 nm. The bond length for (La-O) is equal to 0.322 nm. The result shows that the band length for (Nd-O) is shorter than (La-O). The model for Neodymium Zinc Titanium oxide (NZT) was suggested[25].

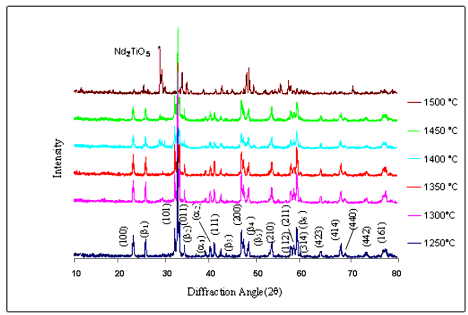

**Figure 10.** X-ray diffraction traces of NZT at 1250-1500 ºC.

### 4.1. Structure of NZT

High-resolution XRD of NZT samples are shown in **Figure 10**. The splitting of fundamental perovskite peaks is particularly evident, for example, in the {110} spacing. The lattice constants of the disordered, double unit cell can be calculated as a= 0.7869 nm, b=0.7564 nm and c= 0.8049 nm. For comparison purposes, the equivalent pseudocubic lattice constant would be a=0.78248 nm which is smaller than $La(Zn_{1/2}Ti_{1/2})O_3$ (LZT) due to smaller size of $Nd^{+3}$ ion.

The reflection marked $\alpha_1$ and $\alpha_2$ in Figure 10 correspond to two sets of {odd,odd,odd} reflection, which are associated with antiphase tilting of oxygen octahedral along at least two axes. The β peaks corresponds to anti parallel displacement of the A-site cation, $Nd^{+3}$.

The Raman spectroscopy of NZT is shown in **Figure 11** and it shows none of characteristics of a typical ordered perovskite. As tolerance factor for NZT (t = 0.916) would be lower than might be expected in a perovskite, it is logical to question weather a perovskite, even if severely tilted, would be stable with this composition.

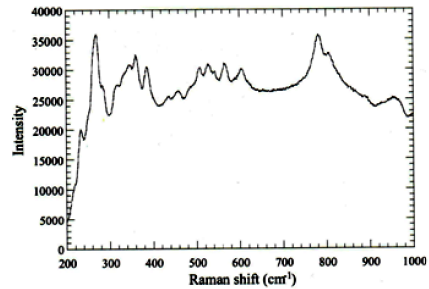

**Figure 11**. The Raman spectrum of NZT.

SEM micrographs of NZT sintered at 1400 ºC for 6 hours are shown in **Figure 12,** pellet is a single-phase, dense ceramic. The NZT has average grain size of $6\,\mu m$. The experimentally, material with different phase must show a different contrast, material with higher atomic number appear brighter. The cracks in Figure 12 are result of mechanical grinding, this occurred during sample preparation.

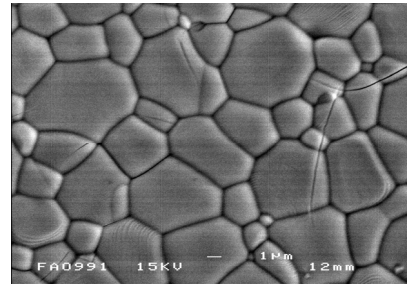

**Figure 12.** Pellet of NZT sintered at 1400 ºC for 6 hours.

### 4.2. Determination of Crystal System

To determine Crystal structure of samples Transmission Electron Microscope (Jeol Jem 2010) was used. Joel JEM is a high resolution analytical microscope fitted with $LaBa_6$ filament operating at acceleration voltage of up to 200 kV and has a lattice resolution of better than 0.2 nm. The equipment is fitted with Oxford Instruments Analyser INCA Energy 300 system.

The crystal structure of samples can be determined by using Kikuchi lines. Because Kikuchi lines exhibit Laue crystal symmetry, they may be used in principle to determine the crystal system to which a phase belongs. However, such determination is generally simpler using alternative x-ray techniques. Some success may be obtained in the electron microscope where large angles of specimen tilt are available.

The Kikuchi line occurs if a Selected Area Diffraction Pattern (SADP) is taken from a single crystal specimen provided the sample is reasonably thick usually it is about half maximum useable penetration and it must have low defect density. Kikuchi line and spot patterns are significance of Single crystal area of the specimen. However, Kikuchi line consists of both pairs of dark and bright lines. By Kikuchi line the orientation relationship between phases can be determine. Kikuchi lines are very useful in orientating the crystal precisely in the TEM. When the sample is tilted, the kikuchi lines move as they were attached to the bottom of the crystal.

The various diffracted spots never move, they just appear and disappear as the angle changes; but Kikuchi line moves with specimen. Following them in a handy way to tilt systematically in order to navigate form one zone axis to another or to find appropriate two beam conditions for example where only one set of planes Bragg diffracts for defect analysis.

Kikuchi lines are only observable if the specimen is thick enough to generate sufficient intensity of scattered electrons.

In a very thin specimen the lines will be too weak to distinguish from the background. As thickness increases, Kikuchi lines and then bands are observed until finally total absorption occurs and nothing is visible. **Figure 13** shows a kikuchi pattern of Neodymium zinc titanium oxide (NZT).

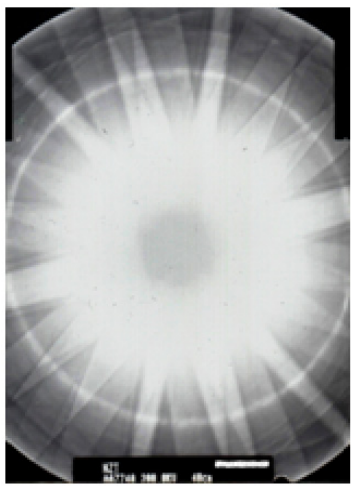

**Figure 13**. Kikuchi Patterns of Single Crystal region of NZT.

Previous data could not be found for NZT analogue, and there was not any information about this material and its crystal structure, as well as its space group. NZT is a material newly discovered during this research work.

After testing the samples by TEM the results indicated that the space group could be one of the following groups: P1, P2$_1$/m, P2/m, I2/a, Pmmn, P4/mmc. β is defined as anti-parallel displacement of $Nd^{+3}$ ions 1/2[ even, even, odd].

At least one axis without (+). No γ (in phase tilting of oxygen octahedra) reflections could be found. All the tests and examinations by TEM, X-Ray, SEM, Raman Spectroscopy, shows that. Finally, the results are showing that monoclinic crystal structure P2$_1$/n is the closest suggested for NZT **(Figure 14)**.

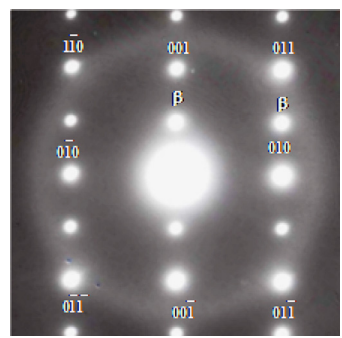

**Figure 14.** [111] Selected area diffraction pattern of the NZT.

**Figure 15**, shows the absence of γ reflections (γ reflections like 1/2$(3\bar{1}\bar{2})$ which indicates in phase tilting of the oxygen octahedra; odd, odd, even, and at least one axis without +). In this system α is equal to antiphase tilting of the oxygen octahedra, for example, 1/2[311]. β is equal to the anti phase displacement of $Nd^{+3}$.

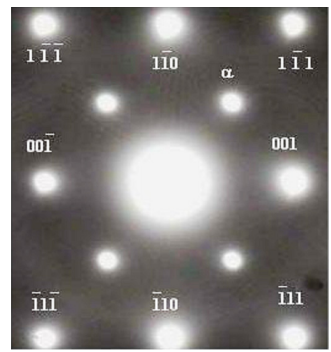

**Figure 15**. [110] Selected is the diffraction pattern of NZT. A-reflections = one, two, or three axes with (-).

The atomic position in NZT is shown in part (a) and (b) of **Figure 16**. The Top view model (a) vertically is based on repetition of Neodymium and Titanium, while horizontally it is Neodymium, Zinc and Titanium on the first top row, at the same time as in the second row; it is Titanium Neodymium and Titanium. This systematically repetition is also shown in other Perovskite crystal structures of Rare Earth Materials. In section (b) of **Figure 16**, the Neodymium is not located on the corner of the cube and there are some distances from the corner, this is the result of the small size of Neodymium which rattles on different crones. The top view and bottom view of the model (b) of NZT shows a hexagonal shape instead of a cube. There is uncertainty about the exact crystal structure of NZT material. This research shows that there are two possibilities to define the exact crystal structure of NZT. The first one is based on a monoclinic crystal structure, and the second is based on a hexagonal.

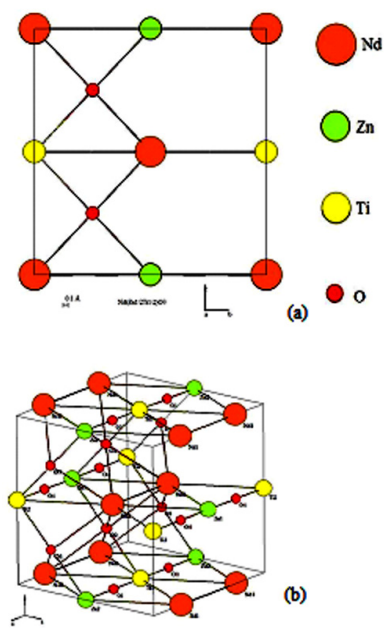

**Figure 16.** The atomic positions in NZT are shown in 100 directions (a). This model is based on monoclinic P2$_1$/n crystal structure (b).

## 5. Conclusions

Single-phase ceramics of Neodymium Zinc Titanium Oxide; have been synthesised in this studies at every sintering temperature 1250-1675 °C. In particular NZT has the temperature coefficient of resonant frequency 47 MK$^{-1}$, Quality factor was 42000 at frequency of 4.33 GHz and relative permittivity 36. The crystal structure of NZT is closer to Crystal system of monoclinic with Bravais Lattice P and space group of P2$_1$/n. Kikuchi line shows that this material has a Single Phase and clearly can be used as filters for the mobile microwave telecommunications.